# Observation of two-dimensional surface solitons in anisotropic waveguide arrays

A. Szameit[1], Y. V. Kartashov[2], F. Dreisow[1], T. Pertsch[1], S. Nolte[1], A. Tünnermann[1,3], and L. Torner[2]

[1]Institute of Applied Physics, Friedrich-Schiller-University Jena, Max-Wien-Platz 1, 07743 Jena, Germany

[2]ICFO-Institut de Ciencies Fotoniques, and Universitat Politecnica de Catalunya, Mediterranean Technology Park, 08860 Castelldefels (Barcelona), Spain

[3]Fraunhofer Institute for Applied Optics and Precision Engineering, Albert-Einstein-Strasse 7, 07745 Jena, Germany

We report on the experimental observation of two-dimensional surface waves localized at the edge or in the corner of femtosecond laser-written waveguide arrays in fused silica. Increasing the power of the input beam allows one to observe a clear transition from a linear diffraction pattern to localized nonlinear surface states, which can exist at the interface only above a certain power threshold. This constitutes the first ever experimental observation of two-dimensional nonlinear surface solitons in optics.

The presence of an interface between different materials can profoundly affect the evolution of nonlinear excitations. Such interface can support stationary surface waves. These were encountered in various areas of physics including solid-state physics [1], near-surface optics [2], plasmas [3], and acoustics [4]. In nonlinear optics, surface waves were under active consideration since 1980. Such waves typically form at the interface when their power exceeds a certain threshold [5-7]. The progress in their experimental observation was severely limited because of unrealistically high power levels required for surface wave excitation at the interfaces of natural materials. However, shallow refractive index modulations accessible in a technologically fabricated waveguide array (or lattice) may facilitate the formation of surface waves at moderate power levels at the edge of semi-infinite arrays as was suggested in Ref. [8]. This has led to the observation of one-dimensional surface solitons in arrays with focusing nonlinearity [9]. Defocusing lattice interfaces are also capable to support surface gap solitons [10-13]. Surface lattice solitons may exist not only in cubic and saturable materials, but also in quadratic [14] and nonlocal [15] media, as well as at the interfaces of complex arrays [16].

Recently, it was suggested that two-dimensional lattice interfaces also can support surface solitons [17-20]. Such systems allow for the existence of complex surface states, such as vortex surface solitons [17], which are not possible in one-dimensional geometries. Two-dimensional array might be used to build complex interfaces exhibiting corners, holes, and engineerable inhomogeneities [20]. In this Letter we report on the first, to the best of our knowledge, experimental observation of two-dimensional surface solitons. Our experiments were conducted at the edge of femtosecond (fs) laser-written waveguide arrays in fused silica.

The fs laser direct writing technique [21] allows fabrication of waveguide arrays along arbitrary paths [22] and with various topologies, such as square [23], hexagonal [24] and circular [25], where multiple waveguides can be specifically excited [26]. Since the nonlinearity of the waveguides is affected by the writing parameters [27], it is possible to tune it for specific purposes, such as excitation of 1D and 2D discrete solitons [28,29]. Using fs laser pulses is a superior technique for the fabrication of large high quality arrays for the investigation of light propagation in the presence of a strong influence of the array boundaries. In this Letter we used such arrays for the observation of nonlinear surface waves which are located both in the central waveguide of the first array row and in the corner waveguide.

To understand the signature of experimental formation of 2D surface waves we studied their properties under ideal continuous wave (CW) illumination. The propagation of laser radiation along the $\xi$ axis in our sample can be described by the nonlinear Schrödinger equation for the dimensionless amplitude of the light field $q$:

$$i\frac{\partial q}{\partial \xi} = -\frac{1}{2}\left(\frac{\partial^2 q}{\partial \eta^2} + \frac{\partial^2 q}{\partial \zeta^2}\right) - q|q|^2 - pR(\eta,\zeta)q, \qquad (1)$$

where the longitudinal $\xi$ and transverse $\eta,\zeta$ coordinates are scaled in terms of the diffraction length and input beam width $r_0$, respectively. We assume a uniform effective focusing nonlinearity inside the sample with $n_2 = 2.7 \times 10^{-20}$ m$^2$/W. The parameter $p$ describes the refractive index modulation inside the sample. For beams with a width $r_0 = 10$ $\mu$m at the wavelength $\lambda = 800$ nm, a value of $p = 3$ corresponds to a real refractive index modulation depth $3 \times 10^{-4}$ in our sample. The function $R(\eta,\zeta) = \sum_{i,k=1,N} Q(\eta - \eta_i, \zeta - \zeta_k)$ describes the refractive index distribution inside the array, where $\eta_i, \zeta_i$ characterize the positions of the individual waveguides separated by distances $w_s$; $N$ is the number of waveguides per dimension, where each waveguide has an elliptical shape $Q(\eta,\zeta) = \exp[-(\eta/w_\eta)^2 - (\zeta/w_\zeta)^2]$ according to the experimental data. We set $N = 5$, $w_s = 5$, $w_\eta = 1.4$, $w_\zeta = 0.4$, which closely resembles the parameters of our $5 \times 5$ fs laser written waveguide array ($14$ $\mu$m $\times$ $4$ $\mu$m widths of the individual waveguides and $50$ $\mu$m separation between them). In Eq. (1) several quantities are conserved, including the energy flow $U = \int\int_{-\infty}^{\infty} |q|^2 \, d\eta d\zeta$.

We searched for ground-state soliton solutions located either in the central waveguide of the first (near-surface) array row or in the corner waveguide, with the form $q = w(\eta,\zeta) \exp(ib\xi)$, where $w(\eta,\zeta)$ is a real function and $b$ is the propagation constant. In the low-power limit surface solitons acquire a multi-peaked structure and expand into the array region (Fig. 1(a)), while with an increase of power the solitons gradually concentrate in the near-surface waveguide (Fig. 1(b)). Despite the limited number of waveguides in our array the shapes of solitons in our system and in a system with a semi-infinite array remain similar almost in the entire domain of existence, except for the limit of almost linear surface modes whose expansion across the array in a finite system is cut off at some value of the propagation constant. This confirms that our system can really be used for the study of surface effects at the interface between a periodic and a uniform medium. We observed that similar shape transformations with increase of power occur for solitons in the corner waveguide of the array (Fig. 1(c)). Decreasing power generates a weakly asymmetric corner surface mode with different rates of amplitude decay along the $\eta$ and $\zeta$ axes, but high-power corner modes (not shown here) have profiles almost identical to those of solitons emerging from the central waveguide. The model (1) predicts that two-dimensional surface waves exist only above a certain minimal value of energy flow similarly to surface waves at the interface of uniform materials [5-7]. Moreover, surface waves can be found only for propagation constants $b$ exceeding a cutoff value $b_{\rm co}$. The energy flow of surface solitons is a nonmonotonic function of $b$ (Fig. 2(a)). The derivative $dU/db$ changes its sign from positive to negative in a narrow region close to the cutoff. At $b \to \infty$ the surface soliton profile gradually approaches that of a Townes soliton of a uniform cubic medium, so that $U(b \to \infty) \approx 5.85$. The dependencies $U(b)$ for solitons located in the corner waveguide and in the central waveguide are very similar to each other, but the minimal power required for the existence of a corner soliton (at $p = 3$ it amounts to $U_{\rm th} = 0.193$) is slightly smaller than that for solitons from the central waveguide ($U_{\rm th} = 0.197$), which is consistent with recent findings [20,30]. The cutoff for soliton existence is a monotonically increasing function of the refractive index modulation depth (Fig. 2(b)), while the minimal energy flow of a surface wave dramatically decreases with $p$. A detailed stability analysis confirmed that surface waves belonging to the regions with $dU/db \leq 0$ are exponentially unstable, while their counterparts with $dU/db > 0$ are stable. The critical value of propagation constant for stabilization of surface waves at the interface of finite waveguide array was found to almost completely coincide with that for an interface of infinite array, indicating that the border of stability domain is not affected by array boundaries. Stable surface waves keep their structure over indefinitely long distances even in the presence of strong broadband input noise.

For the experimental investigation of the nonlinear surface waves we fabricated a $5 \times 5$ array with a waveguide separation of $50\,\mu$m and a length of 74.4 mm using a Ti:Sapphire laser system (RegA/Mira, Coherent Inc.) which exhibits a repetition rate of 100 kHz, a pulse duration of about 150 fs and $0.3\,\mu$J pulse energy at a laser wavelength of 800 nm. The beam was focused into a polished fused-silica sample by a $20 \times$ microscope objective with a numerical aperture of 0.45 whereas the used fused silica (Suprasil 311, Haereaus) is of highest quality concerning inhomogeneity and impurities inside the sample (total cross section $<\ 0.01\ {\rm mm}^2/100\ {\rm cm}^3$). The writing velocity was $1250\,\mu$m/s, performed by a high precision positioning system (ALS 130, Aerotech). These writing parameters have been already used several times [27-29] in different fused silica samples proving the high reproducibility of the experimental data. The resulting index changes were determined by measuring the near-field profile at a wavelength of 800 nm, and solving the Helmholtz equation [31] which yields $4 \times 14\,\mu{\rm m}^2$ size of a single waveguide. The transmission losses of a single waveguide, measured by a cut-back method, were $< 0.4\,\mathrm{dB/cm}$. In order to avoid damage of the device when exciting with high power laser pulses, the waveguides are buried $0.5\,\mathrm{mm}$ away from the incoupling facet. This reduces the applied fluence at the sample surface which has a significantly lower damage threshold than the bulk material. Therefore, pulses at a substantially higher peak power can be coupled into the waveguides. For the excitation of the nonlinear surface waves we used a Ti:Sapphire CPA laser system (Spitfire, Spectra-Physics) with a pulse duration of about $150\,\mathrm{fs}$ and a repetition rate of $1\,\mathrm{kHz}$ at $800\,\mathrm{nm}$. The light was coupled either into the central or corner waveguides with a $4\times$ microscope objective $(\mathrm{NA} = 0.1)$, coupled out by a $10\times$ objective $(\mathrm{NA} = 0.25)$, and projected onto a CCD-camera. In Figs. 3 and 4 the time-integrated experimental data are shown, which are consistent with the results of simulations using model (1) with the input conditions $q|_{\xi=0} = A\exp(-\eta^2 - \zeta^2)$ describing the excitation of the central waveguide of the first array row or corner waveguide by Gaussian beams with various amplitudes $A$ ($A = 0.3$ in panels (a), 0.37 in (b), and 0.6 in (c)). At a below-threshold input peak power (1.2 MW) the experimental pattern broadens due to discrete diffraction and almost all of the power in the excited waveguides has been coupled into the adjacent ones (Fig. 3(a), 4(a)). This is consistent with the beam dynamics governed by the model (1) with input beam power far below the CW threshold ($U_{\mathrm{th}} \approx 0.197$) for surface soliton formation, so that no soliton forms. Figure 4(a) shows linear coupling from the corner waveguide, which is only marginal influenced by the anisotropy of the waveguides, since in our case the waveguide separations are large, so that the coupling is only weakly anisotropic [25]. It is a welcome fact that despite the large vertical array dimensions the upper and lower waveguides are identical, so that the obtained results are independent on the excited corner. At a 1.8 MW peak power a slightly localized intermediate state is observed (left panels in Figs. 3(b) and 4(b)), that is consistent with the pattern obtained for input peak powers slightly exceeding the CW threshold for surface wave formation (right panels). Notice that the experimentally observed threshold power for soliton formation with pulsed light is expected to slightly exceed the theoretical value calculated for CW illumination, since even when the peak power exceeds the soliton threshold the pulse wings still diffract yielding a broaden time-integrated pattern. The impact of pulse wings on the output patterns becomes negligible with increasing input peak power and thus unambiguously localized surface states are readily observed for high enough input powers (Figs. 3(c) and 4(c)). Such observed patterns are consistent with theoretical predictions for CW Gaussian beams with powers substantially exceeding the CW threshold. Under the assumption that most of the power of the input Gaussian beam couples into the stationary surface soliton one may conclude that the beams in panels (b) and (c) would correspond to CW solitons corresponding to propagation constant values $b \approx 0.6$ and $b \approx 0.65$, respectively, both belonging to the stable branch of the $U(b)$ curve. Note, however, that experiments are conducted with pulsed pump light, therefore such estimates are only to confirm consistency of the observations with soliton formation around the pulse peak. Due to the high accuracy of the writing process the array exhibits very homogeneous coupling between the waveguides in $\eta$ and $\zeta$ directions and sharply defined edges of the waveguide array. This ensures the absence of scattering or noticeable statistical distortions of the output pattern caused by small waveguide displacements or inhomogeneities. Thus, the fs laser writing technique is ideally suited for studying surface phenomena.

Summarizing, we demonstrated experimentally that the nonlinear interface between an array of fs laser written waveguides and the uniform medium can support different types of stable two-dimensional surface waves. Our observations prove the high potential of the laser writing technique for the creation of complex refractive index landscapes, including surfaces with desirable geometry. Nonlinear discrete surface waves might find potential applications in beam routing schemes and in surface characterization and sensing; they also motivate potential analogies with other areas of physics.

This work has been supported by the Deutsche Forschungsgemeinschaft (Research Unit 532 "Nonlinear spatial-temporal dynamics in dissipative and discrete optical systems"), the German Federal Ministry of Education and Research (Innoregio, 03ZIK051), the Jenoptik AG, and the Government of Spain through grants TEC2005-07815, Accion Integrada HA2005-0107, and the Ramon-y-Cajal program.

Note added: We note that observation of surface solitons at optically induced lattice interfaces has been recently reported in Ref. [32].

# References

1. *Nonlinear waves in solid state physics*/ Ed. by A. D. Boardman, M. Bertolotti, and T. Twardowski, NATO ASI **247**, Plenum Press, New York (1989).
2. *Near-Field Optics and Surface Plasmon Polaritons*/ Ed. by S. Kawata, Springer, Berlin (2001).
3. Y. M. Aliev, H. Schluter, and A. Shivarova, *Guided-wave-produced plasmas* (Springer, Berlin, 2000).
4. S. V. Biryukov, Y. V. Gulyaev, V. V. Krylov, V. P. Plessky, *Surface acoustic waves in inhomogeneous media* (Springer, Berlin, 1995).
5. *Nonlinear surface electromagnetic phenomena*/ Ed. by H. E. Ponath and G. I. Stegeman, North Holland, Amsterdam (1991).
6. D. Mihalache, M. Bertolotti, and C. Sibilia, "Nonlinear wave propagation in planar structures," Progr. Opt. **27**, 229 (1989).
7. A. A. Maradudin, in *Optical and Acoustic Waves in Solids - Modern Topics*, M. Borissov Ed., (World Scientific, Singapore, 1983), 72.
8. K. G. Makris *et al.*, Opt. Lett. **30**, 2466 (2005).
9. S. Suntsov *et al.*, Phys. Rev. Lett. **96**, 063901 (2006).
10. Y. V. Kartashov, V. A. Vysloukh, and L. Torner, Phys. Rev. Lett. **96**, 073901 (2006).
11. M. I. Molina, R. A. Vicencio, and Y. S. Kivshar, Opt. Lett. **31**, 1693 (2006).
12. C. R. Rosberg *et al.*, Phys. Rev. Lett. **97**, 083901 (2006).
13. E. Smirnov *et al.*, Opt. Lett. **31**, 2338 (2006).
14. G. A. Siviloglou *et al.*, Opt. Express **14**, 5508 (2006).
15. Y. V. Kartashov, L. Torner, and V. A. Vysloukh, Opt. Lett. **31**, 2595 (2006).
16. M. I. Molina *et al.*, Opt. Lett. **31**, 2332 (2006).
17. Y. V. Kartashov *et al.*, Opt. Express **14**, 4049 (2006).
18. Y. V. Kartashov and L. Torner, Opt. Lett. **31**, 2172 (2006).
19. Y. V. Kartashov *et al.*, Opt. Lett. **31**, 2329 (2006).
20. K. G. Makris *et al.*, Opt. Lett. **31**, 2774 (2006).
21. K. Itoh *et al.*, MRS Bulletin **31,** 620 (2006).

22. S. Nolte *et al.*, Appl. Phys. A **77**, 109 (2003).
23. T. Pertsch *et al.*, Opt. Lett. **29**, 468 (2004).
24. A. Szameit *et al.*, Appl. Phys. B **82**, 507 (2006).
25. A. Szameit *et al.*, Opt. Express 15, 1579 (2007).
26. A. Szameit *et al.*, Appl. Phys. B 87, 17 (2007).
27. D. Bloemer *et al.*, Opt. Express **14**, 2151 (2006).
28. A. Szameit *et al.*, Opt. Express **13**, 10552 (2005).
29. A. Szameit *et al.*, Opt. Express **14**, 6055 (2006).
30. A. Vicencio *et al.*, "Discrete surface solitons in two-dimensional anisotropic photonic lattices," Phys. Lett. A (in press).
31. I. Mansour and F. Caccavale, J. Lightwave Technol. **14**, 423 (1996).
32. X. Wang et al., Phys. Rev. Lett. **98**, 123903 (2007).

# Figure captions

Figure 1.      Profiles of surface solitons located in the central waveguide of the first array row at $b = 0.57$ (a) and $0.72$ (b). Profile of a surface soliton located in the corner waveguide at $b = 0.57$ (c). In all cases $p = 3$.

Figure 2.      (a) Energy flow versus propagation constant for a surface soliton located in the central waveguide of first array row for $p = 3$. Points marked by circles correspond to profiles shown in Figs. 1(a) and 1(b). Notice, that the minimal energy flow $U_{\text{th}} = 0.197$ for existence of soliton from the central waveguide is slightly higher than minimal energy $U_{\text{th}} = 0.193$ of corner soliton. (b) Cutoff versus refractive index modulation depth for soliton from central waveguide.

Figure 3 (color online).      Excitation of a surface wave in the central waveguide of the first array row. Left column – experiment, right column – simulation. Input peak power is 1.2 MW (a), 1.8 MW (b), and 4.8 MW (c). Dashed lines indicate interface position.

Figure 4 (color online).      Excitation of a surface wave in the corner waveguide of the array. Left column – experiment, right column – simulation. Input peak power is 1.2 MW (a), 1.8 MW (b), and 4.8 MW (c).

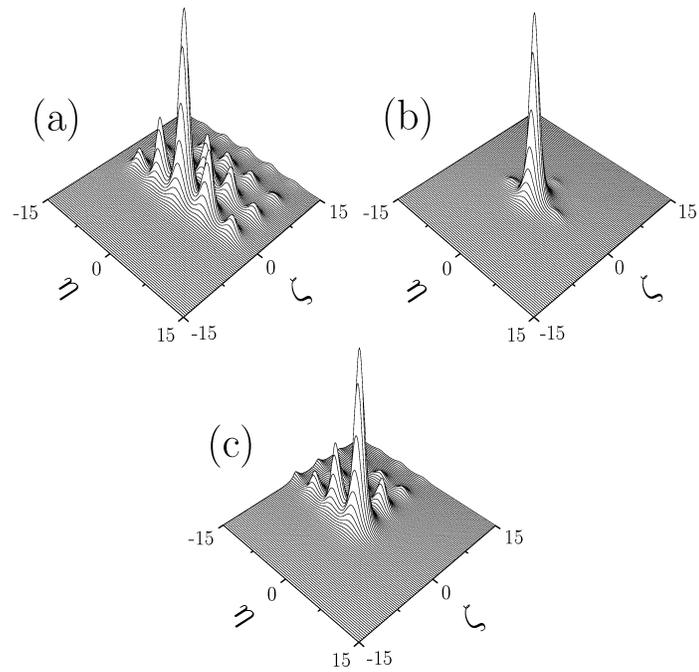

Figure 1. Profiles of surface solitons located in the central waveguide of the first array row at $b = 0.57$ (a) and $0.72$ (b). Profile of a surface soliton located in the corner waveguide at $b = 0.57$ (c). In all cases $p = 3$.

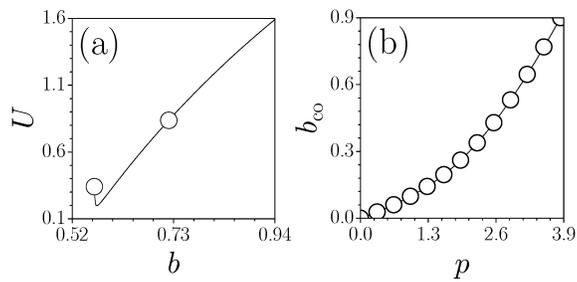

Figure 2.  (a) Energy flow versus propagation constant for a surface soliton located in the central waveguide of first array row for $p = 3$. Points marked by circles correspond to profiles shown in Figs. 1(a) and 1(b). Notice, that the minimal energy flow $U_{\text{th}} = 0.197$ for existence of soliton from the central waveguide is slightly higher than minimal energy $U_{\text{th}} = 0.193$ of corner soliton. (b) Cutoff versus refractive index modulation depth for soliton from central waveguide.

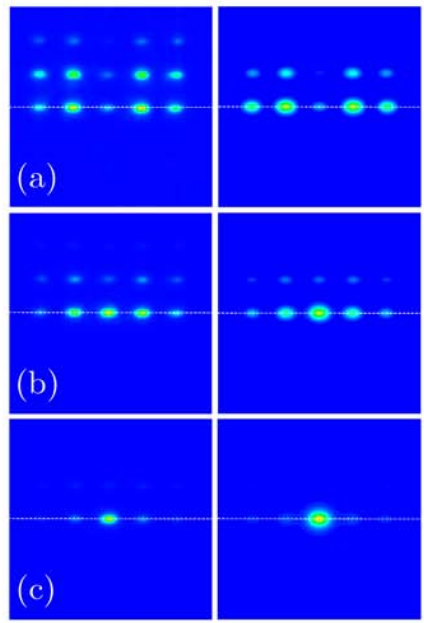

Figure 3 (color online). Excitation of a surface wave in the central waveguide of the first array row. Left column – experiment, right column – simulation. Input peak power is 1.2 MW (a), 1.8 MW (b), and 4.8 MW (c). Dashed lines indicate interface position.

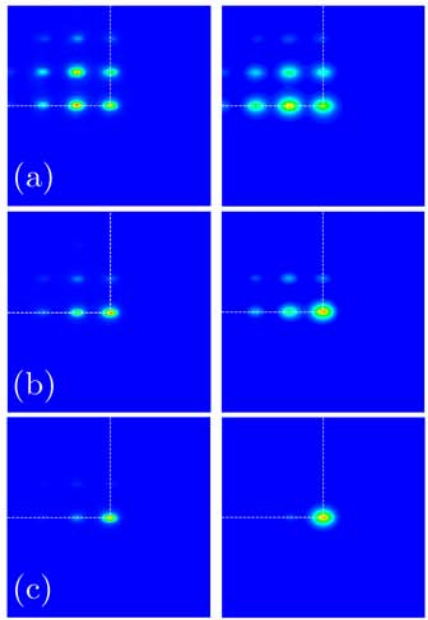

Figure 4 (color online). Excitation of a surface wave in the corner waveguide of the array. Left column – experiment, right column – simulation. Input peak power is 1.2 MW (a), 1.8 MW (b), and 4.8 MW (c).